\documentstyle[graphicx, 12pt]{article}
\begin{document}

\small
\hoffset=-1truecm
\voffset=-2truecm
\title{\bf The description of phase transition of Bardeen black hole in the Ehrenfest scheme}
\author{Jingyun Man \hspace {1cm} Hongbo Cheng\footnote {E-mail address: hbcheng@ecust.edu.cn}\\
Department of Physics, East China University of Science and
Technology,\\ Shanghai 200237, China\\
The Shanghai Key Laboratory of Astrophysics, Shanghai 200234,
China}

\date{\today}
\maketitle

\begin{abstract}
The phase transition of a Bardeen black hole is studied by
considering Ehrenfest's equations. The thermodynamic variables
such as entropy, potential and heat capacity are calculated from
the first law of thermodynamics for black holes. That no
discontinuity in entropy and potential of the black holes means
that the first order phase transition will not generate for the
Bardeen black holes. However, the divergence of heat capacity at
constant potential and satisfaction of the Ehrenfest's equations
indicates that the second order phase transition of Bardeen black
hole will appear.
\end{abstract}
\vspace{7cm} \hspace{1cm} PACS number(s): 04.70Bw, 14.80.Hv

\newpage

\noindent \textbf{1.\hspace{0.4cm}Introduction}

It is important that black holes can be treated as thermodynamic
objects with physical temperature and entropy. Bekenstein has
claimed that the black holes should have temperature
which is not equal to zero and the intrinsic entropy of a black
hole is proportional to the area of event horizon [1-3]. It has
been shown that a black hole has a thermal radiation with the
temperature due to its surface gravity through studying the
quantum mechanics of scalar particles around a black hole by
Hawking et.al [4-6].

Bradeen model describes a "regular" spacetime without a
singularity but with a horizon [7-9]. The Bardeen black hole and
several other regular black holes can be presented as the exact
solutions of nonlinear electrodynamics coupled to Einstein gravity
[10-13]. Subsequently, the gravitational lensing of Bardeen black
holes was studied in Ref.[14]. The gravitational and
electromagnetic stability of Bardeen black hole was explored [15].
The quasinormal modes of the Bardeen black hole were also
discussed [16]. Sharif and Javed have noticed the thermodynamical
quantities as temperature and entropy for a Bardeen black hole by
means of a quantum tunneling approach over semiclassical
approximations [17]. The effects of space noncommutativity on the
thermodynamics of a Bardeen black hole was also examined in
Ref.[18]. However, we notice that the entropy of the Bardeen black
hole is a function of not only the event horizon radius of the
black hole but also the charge of magnetic monopole swallowed by
the black hole. During the calculation of electric potential, we
set the entropy as a constant and obtain an electric potential
which is different from that in Ref.[17] and Ref.[18].

The Ehrenfest scheme is a powerful tool to describe the phase
transition of black holes [19]. In the context of this scheme,
Banerjee et. al confirmed that the phase transitions from
liquid-vapour systems to Reissner-Nordstrom-AdS or Kerr-AdS black
holes belong to the second order [20, 21]. We also discuss the
phase transition of a black hole with conformal anomaly with the
help of Ehrenfest's equation [22]. In this process the basic
classical thermodynamic method is transplanted to study the order
of phase transition in black holes. If the Gibbs potential of the
system continues while the first order derivative of Gibbs
potential is discontinuous, the phase transition is so-called
first order. The so-called second order phase transition can be
thought as that the Gibbs potential of the system and its first
order derivative is continuous while the second order derivative
of the potential is not continuous. The higher order transitions
can be studied in the way on the analogy of this.

In this paper, we would like to study the phase transitions of
Bardeen black holes in virtue of the Ehrenfest's scheme with
analytical approximation to the critical points of phase
transitions. First we conform to Hawking temperature and the first
law of thermodynamics, then obtain the black hole entropy with the
logarithmic term and the characteristic electric potential which
is distinguished from the results before like Ref. [17,18].
Absence of any discontinuity in entropy and potential of the black
hole as the functions with horizon radius and infinite divergences
in heat capacity at constant potential indicates that the higher
order phase transition of Bardeen black hole occurs. The explicit
relation of critical horizon radius and charge is given. We then
show the first and second Ehrenfest's equations are established at
the critical phase transition point. The phase transition of
second order will be confirmed. The paper is organized as follows.
In section 2, we recapitulate the Bardeen black hole. The explicit
expressions of thermodynamic quantities like temperature, entropy,
electric potential and heat capacity of the black hole are given,
and we discuss the critical points of phase transition. In section
3, we prove that the phase transition of a Bardeen black hole is
second order with the help of Ehrenfest equation. Finally, the
discussion and conclusions are emphasized.

\vspace{0.8cm} \noindent \textbf{2.\hspace{0.4cm}The thermodynamic
quantities of Bardeen black holes}

At first we will give a brief introduction to the Bardeen black
hole as a regular solution of Einstein equations coupled to a
nonlinear electrodynamics which provide a monopole charge $q$
[10]. The action was proposed as,

\begin{equation}
S=\int dv \left(\frac{1}{16\pi}R-\frac{1}{4\pi }{\mathcal{L}}(F)\right)  ,
\end{equation}

\noindent where $R$ is the scalar curvative, and $\mathcal{L}$ is
the Lagrangian, a function of $F=\frac{1}{4}F_{\mu\nu}
F^{\mu\nu}$, and $F_{\mu\nu}=2\nabla_{[\mu}A_{\nu]}$ is the
electromagnetic strength. $\mathcal{L}$ as the nonlinear
electrodynamics source is denoted as,

\begin{equation}
{\mathcal{L}}(F)=\frac{3M}{|q|q^{2}}\left(\frac{\sqrt{2q^{2}F}}
{1+\sqrt{2q^{2}F}}\right)^{\frac{5}{2}} .
\end{equation}

\noindent Here $q$ and $M$ are magnetic charge and the mass of the
magnetic monopole respectively. The Einstein-nonlinear
electrodynamic field equations are

\begin{equation}
G_{\mu}^{\nu}=2\left(\frac{\partial{\mathcal{L}}}{\partial
F}F_{\mu\lambda}F^{\nu\lambda}-\delta_{u}^{v}\mathcal{L}\right) ,
\end{equation}

\begin{equation}
\nabla_{\mu}\left(\frac{\partial{\mathcal{L}}}{\partial
F}F^{\alpha\mu}\right)=0 .
\end{equation}

The Bardeen black hole solution with a static and spherically
symmetric configuration is provided by [7],

\begin{equation}
ds^{2}=-f(r)dt^{2}+\frac{dr^{2}}{f(r)}+r^{2}(d\theta^{2}+\sin^{2}\theta
d\varphi^{2}) ,
\end{equation}

\noindent where

\begin{equation}
f(r)=1-\frac{2Mr^{2}}{(r^{2}+q^{2})^{\frac{3}{2}}} .
\end{equation}

\noindent Note for $q=0$, it recovers to the well-known
Schwarzschild metric. Contrary to traditional black hole spacetime
which are expected to have horizons and singularities covered by
the horizons, the Bardeen model has a horizon but no singularity.
We find the event horizon $r_{H}$ of the black hole as a solution
of $f(r)=0$. The total mass of the Bardeen black hole can be
presented as

\begin{equation}
M=\frac{(r_{H}^{2}+q^{2})^{\frac{3}{2}}}{2r_{H}^{2}} .
\end{equation}

\noindent The relation between the mass and the event horizon
radius is drawn in Figure 1. There is a minimum
$M_{0}=\frac{3\sqrt{3}}{4}q$ for $r_{H}=\sqrt{2}q$.

We shall calculate the thermodynamic variables of the Bardeen
black hole in order to discuss the phase transition happened in
this kind of black hole. We start with Hawking temperature defined
as [4],

\begin{equation}
T_{H}(q,r_{H})=-\frac{1}{4\pi}\left[\sqrt{-g^{tt}g^{rr}}\frac{dg_{tt}}
{dr}\right]_{r=r_{H}}\nonumber\\
=\frac{1}{4\pi}\frac{r_{H}^{2}-2q^{2}}{r_{H}(r_{H}^{2}+q^{2})}\hspace{2cm}
\end{equation}

\noindent shown in Figure 2. From Figure 2, it is clear that the
temperature of the Bardeen black hole cannot exceed the threshold.
The maximum temperature $T_{0}$ of an extremal configuration with
radius $r_{0}$ are obtained,

\begin{equation}
T_{0}=\frac{(r_{0}^{2}-2q^{2})}{4\pi r_{0} (r_{0}^{2}+q^{2})} ,
\end{equation}

\begin{equation}
r_{0}=\frac{\sqrt{14+2\sqrt{57}}}{2} q .
\end{equation}

The first law of thermodynamics for black holes can be written as
[1-3, 20],

\begin{equation}
dM=T_{H}dS+\phi dq ,
\end{equation}

\noindent where $S$ is the modified entropy and $\phi $ is the
electric potential. Here the semiclassical entropy can be
expressed as [20],

\begin{equation}
S=\int\left(\frac{1}{T}\frac{\partial M}{\partial r_{H}}\right)_{q} dr_{H}
\nonumber=\frac{\pi}{r_{H}}(r_{H}^{2}-2q^{2})\sqrt{r_{H}^{2}+q^{2}}
+3\pi q^{2}ln\left(r_{H}+\sqrt{r_{H}^{2}+q^{2}}\right) .
\end{equation}

\noindent The subscript $q$ requires that the charge has to be
treated as an invariant during calculating the integration.
According to Eq. (7), the total mass of the black hole involves
the charge $q$. It recovers to area entrpy $\pi r_{H}^{2}$, if
charge $q$ reduces to zero. In Figure 3, there is a continuous
curve which shows the relation between the entropy and the event
horizon radius of the Bardeen black.

According to the first law of thermodynamics, the analytic
expression of electric potential is written as [20],

\begin{eqnarray}
\phi=\left(\frac{\partial M}{\partial q}\right)_{S}\hspace{9.5cm}\nonumber\\
=\frac{3q}{4r_{H}^{2}(r_{H}^{2}+q^{2})(r_{H}+\sqrt{r_{H}^{2}+q^{2}})}
[3r_{H}^{4}+3r_{H}^{3}\sqrt{r_{H}^{2}+q^2}+4q^{2}r_{H}^{2}-2q^{4}\nonumber\\
-2r_{H}(r_{H}^{2}-2q^{2})(r_{H}+\sqrt{r_{H}^{2}+q^{2}})ln(r_{H}+\sqrt{r_{H}^{2}+q^{2}})] .
\end{eqnarray}

\noindent The expression of potential is more cumbersome due to
the tangle among the entropy, event horizon radius and charge.
Since the entropy is not simply proportional to the area of event
horizon, it no longer works here for choosing horizon radius
$r_{H}$ as an invariant instead of entropy $S$. The trick we used
here is treating horizon radius as a function of charge. We plot
the electric potential dependent on black hole radius for $q=0.9$
in Figure 4. Neither entropy and potential have a discontinuity upon
the behavior of horizon radius, so the first order phase
transition of the Bardeen black hole is ruled out.

In addition, the Gibbs free energy for Bardeen black hole is defined as [20],

\begin{equation}
G=M-TS-\phi q.
\end{equation}

\noindent With thermodynamic quantities Eq. (7), Eq. (8), Eq. (12)
and Eq. (13), the free energy $G$ can be expressed as a function
of event horizon radius $r_{H}$ and magnetic charge $q$, showed as
a continuous curve in Figure 5. It can be easily checked that the
non-negative definiteness $T_{H}$ give a strong constraint on the
horizon radius  $r_{H}\geq \sqrt{2} q$.

The same procedure [19-21] may be easily adapted to obtain heat
capacity at constant potential,

\begin{equation}
C_{\phi}=\left(T\frac{\partial S}{\partial T}\right)_{\phi}
=\frac{B(r_{H},q)}{A(r_{H},q)} ,
\end{equation}

\noindent where the function $A(r_{H},q)$ is

\begin{eqnarray}
A(r_{H},q)=-2\sqrt{r_{H}^{2}+q^{2}}(-3r_{H}^{6}-14 r_{H}^{4}q^{2}
+3 r_{H}^{2}q^{4}+2q^{6})
-r_{H} (2q^{6}-5 r_{H}^{2}q^{4}-31 r_{H}^{4} q^{2}-6r_{H}^{6})\nonumber\\
+2ln(r_{H}+\sqrt{r_{H}^{2}+q^{2}})(-r_{H}^{4}+7q^{2}r_{H}^{2}
+2q^{4})r_{H}(r_{H}+\sqrt{r_{H}^{2}+q^{2}})^{2}\nonumber\\ ,
\end{eqnarray}

\noindent and $B(r_{H},q)$ is

\begin{eqnarray}
B(r_{H},q)=\frac{-\pi}{r_{H}+\sqrt{r_{H}^{2}+q^{2}}}
[12(-r_{H}^{4}+7q^{2}r_{H}^{2}+2q^{4})ln(r_{H}
+\sqrt{r_{H}^{2}+q^{2}})^{2}r_{H}^{2}q^{2}\nonumber\\
\times(4r_{H}^{3}+3r_{H}q^{2}+(4r_{H}^{2}+q^{2})\sqrt{r_{H}^{2}+q^{2}})\nonumber\\
-4r_{H}ln(r_{H}+\sqrt{r_{H}^{2}+q^{2}})(4r_{H}^{10}-55r_{H}^{8}q^{2}
-56r_{H}^{6}q^{4}+70r_{H}^{4}q^{6}+59r_{H}^{2}q^{8}\nonumber\\
+10q^{10}+r_{H}\sqrt{r_{H}^{2}+q^{2}}(4r_{H}^{8}-57r_{H}^{6}q^{2}
-27r_{H}^{4}q^{4}+76r_{H}^{2}q^{6}+24q^{8}))\nonumber\\
+\sqrt{r_{H}^{2}+q^{2}}(-2r_{H}^{8}q^{2}-59r_{H}^{6}q^{4}
+82r_{H}^{2}q^{8}+51r_{H}^{4}q^{6}+16q^{10}+24r_{H}^{10})\nonumber\\
+r_{H}(24r_{H}^{10}-63r_{H}^{6}q^{4}+21r_{H}^{4}q^{6}+36q^{10}
+124r_{H}^{2}q^{8}+10r_{H}^{8}q^{2}) ] .
\end{eqnarray}

\noindent Heat capacity diverges if the denominator $A(r_{H}, q)$
is equal to zero. Then we find that the critical radius point
$r_{c}$ is a root of equation $A(r_{c}, q)=0$. For charge $q=0.9$,
we get a reasonable critical horizon radius $r_{c}=4.877810899$
which is larger than $\sqrt{2} q$. Further we substitute $r_{c}$
into equation (8) to find the critical temperature $T_{c}$ where
some discontinuities happen in the heat capacity at constant
potential while the entropy and potential of the black hole is
continuous. So the phase transition of Bardeen black hole must
higher than the first order. We plot heat capacity at constant
potential expressed with equations (15), (16) and (17) in Figure 6
and Figure 7 as functions of horizon radius and temperature
respectively. In Figure 6, the heat capacity at constant potential
flips from positive infinity to negative infinity along with
increase of event horizon radius. A thermodynamically stable case
requires a positive heat capacity. Noticing that horizon radius
has to be larger than $\sqrt{2} q$,  we find the range of horizon
radius $r_{m}<r_{H}<r_{c}$ for the heat capacity at constant
potentical greater than zero. Here $r_{m}$ is the maximum root of
equation $C_{\phi}=0$. It demonstrates that nearby the critical
point the Bardeen black hole transform from an unstable phase with
larger horizon radius $r_{H}>r_{c}$ where heat capacity in this
situation is negative into a stable phase with smaller black hole
radius $r_{m}<r_{H}<r{c}$ where $C_{\phi}$ is positive. The larger
Bardeen black holes can not survive. From Figure 1 and Figure 2, a
Bardeen black hole in an unstable phase has larger mass
$M>M|_{r_{H}=r_{c}}$ and lower temperature $T_{H}<T_{c}$. It can
not keep this phase for long time and eventually transmutes
through critical point into a stable black hole with smaller mass
$M_{0}<M<M|_{r_{H}=r_{c}}$ but higher temperature
$T_{0}>T_{H}>T_{c}$. We should pay attention to the critical
temperature $T_{c}$ in Figure 7 less than the maximal value
$T_{0}$, which means that this phase transition can happen in an
appropriate temperature. Furthermore, no discontinuity for heat
capacity appears if $q=0$ corresponding to general Schwarzchild
black hole spacetime.

\vspace{0.8cm} \noindent \textbf{3.\hspace{0.4cm}The description
of phase tansition in Ehrenfest relation}

In this section, we will examine whether the phase transition of a
Bardeen black hole is second order by adopting Ehrenfest equation.
It has been showed that the special entropy and electric potential
are continuous, so through the critical point the right-hand-side
of Clausius-Clapeyron equation $-\frac{d\phi}{dT}=\frac{\Delta
S}{\Delta q}$ become indeterminate form $\frac{0}{0}$. Take
advantage of L'Hospital's rule, then two Ehrenfest's equations for
Bardeen spacetime can be easily found [19-21],

\begin{equation}
-\left(\frac{\partial \phi}{\partial T_{H}}\right)_{q}
=\frac{\alpha _{2}-\alpha_{1}}{\kappa_{2}-\kappa_{1}} ,
\end{equation}

\begin{equation}
-\left(\frac{\partial \phi}{\partial T_{H}}\right)_{S}
=\frac{C_{\phi 2}-C_{\phi 1}}{Tq(\alpha _{2}-\alpha_{1})} ,
\end{equation}

\noindent where $\alpha$ is the charge growth coefficient analogy
to volume expansion coefficient, and $\kappa$ is the isothermal
compressibility defined as $\alpha=\frac{1}{q}\left(\frac{\partial
q}{\partial T_{H}}\right)_{\phi}$ and
$\kappa=\frac{1}{q}\left(\frac{\partial
q}{\partial\phi}\right)_{T_{H}}$. With thermodynamic quantities of
this system including magnetic charge $q$, Hawking temperature
$T_{H}$, and electric potential $\phi$, the specific expressions
of charge growth coefficient and isothermal compressibility are
given by

\begin{equation}
\alpha=\frac{C(r_{H},q)}{A(r_{H},q)} ,
\end{equation}

\begin{equation}
\kappa=\frac{D(r_{H},q)}{A(r_{H},q)} ,
\end{equation}

\noindent where the function $C(r_{H},q)$ is given by,
\begin{eqnarray}
C(r_{H},q)=\frac{4\pi r_{H}(r_{H}^{2}+q^{2})}{r_{H}^{2}-2q^{2}}
[2(r_{H}^{4}-7q^{2}r_{H}^{2}-2q^{4}) ln(r_{H}+\sqrt{r_{H}^{2}+q^{2}})
r_{H} (r_{H}+\sqrt{r_{H}^{2}+q^{2}})^{2}\nonumber\\ -r_{H}(5r_{H}^{4}q^{2}
-17r_{H}^{2}q^{4}+10r_{H}^{6}-6q^{6}) +2(2q^{4}+5r_{H}^{2}q^{2}
-5r_{H}^{4})(r_{H}^{2}+q^{2})^{\frac{3}{2}}  ] ,
\end{eqnarray}

\noindent and the function $D(r_{H},q)$ is,
\begin{eqnarray}
D(r_{H},q)=\frac{4(r_{H}+\sqrt{r_{H}^{2}+q^{2}})^{2}(r_{H}^{2}
+q^{2})(r_{H}^{4}-7q^{2}r_{H}^{2}-2q^{4})r_{H}^{2}}
{3q(r_{H}^{2}-2q^{2}) } .
\end{eqnarray}

\noindent They diverges while the black hole as a thermodynamic
object approaches the critical point. The Figure 7 and Figure 8
describe the behavior of charge growth coefficient and isothermal
compressibility depending on the event horizon radius of Bardeen
black hole. Some discontinuities also emerge to exhibit the higher
order phase transition in a Bardeen black hole. One can see from
equations (15), (20) and (21), all of the heat capacity at
constant potential, charge growth coefficient and volume expassion
coefficient have the same denominator. These three thermodynamical
quantities diverge at the same radius $r_{c}$ although some
discontinuities at other points also appear. Therefore, the
critical horizon radius is still our phase transition position.

Firstly we check the left-hand-side of Ehrenfest's equations (18),
(19) at critical point in order to ensure the second order phase
transition is able to occur. Combine the electric potential (13),
and Hawking temperature (8) and take magnetic charge or
entropy (12) as constants, the left-hand-side of equation (18) and
(19) then be found

\begin{equation}
-\left(\frac{\partial \phi}{\partial T_{H}}\right)_{q}
|_{r_{H}=r_{c}}=-\left(\frac{\partial \phi}{\partial T_{H}}
\right)_{S}|_{r_{H}=r_{c}} =-6\pi q .
\end{equation}

Secondly we study the right-hand-side of equations (18), (19) at
critical position. By regarding $q$ invariable, $B(r_{H})$ from
heat capacity at a constant potential, $C(r_{H})$ from the charge
growth coefficient and $D(r_{H})$ from the isothermal
compressibility are well behavior, and all of $C_{\phi}$, $\alpha$
and $\kappa$ have the same denominator $A(r_{H})$. We choose two
arbitrary points $r_{1}$ and $r_{2}$ close to the phase transition
point $r_{c}$, one get $B(r_{1})=B(r_{2})=B(r_{c})$,
$C(r_{1})=C(r_{2})=C(r_{c})$, and $D(r_{1})=D(r_{2})=D(r_{c})$,
and they are all not equal to zero. However, this approximation
method to critical point is not suitable for the denominator term
because $A(r_{c})=0$. The right-hand-side of Ehrenfest's equations
become

\begin{equation}
\frac{\alpha _{2}-\alpha_{1}}{\kappa_{2}-\kappa_{1}}
|_{r_{c}}=\frac{C(r_{c})}{D{r_{c}}}=-6\pi q ,
\end{equation}

\noindent and
\begin{equation}
\frac{C_{\phi 2}-C_{\phi 1}}{Tq(\alpha _{2}-\alpha_{1})}
|_{r_{c}}=\frac{B(r_{c})}{T_{H}qC_{r_{c}}} |_{r_{c}}=-6\pi q .
\end{equation}

\noindent Divergences in heat capacity, charge growth coefficient
and isothermal compressibility are canceled in both of Eq.(16) and
Eq.(17). From Eq.(24), Eq.(25) and Eq.(26), it is evident that the
two Ehrenfest's equations are established at the critical point.
Furthermore the first and second equations are identical to each
other at the critical point $r_{c}$. So far we analytically prove
that the phase transition of a Bardeen black hole belongs to a
second order transition by means of Ehrenfest scheme.

\vspace{0.8cm} \noindent \textbf{4.\hspace{0.4cm}Discussion}

In this paper, with the help of Ehrenfest schemes,
we show analytically that the second order phase transition
will appear for the Bardeen black holes.
According to the first law of thermodynamics
of black holes, we obtain the electric potential, heat capacity at
constant potential, charge growth coefficient and isothermal
compressibility for a regular Bardeen black hole. The entropy and
potential of black hole as functions of event horizon radius have
no discontinuity while the relation between heat capacity and
horizon radius approaches to the infinity at the critical radius
regarded as the signature of phase transition. Close to the
critical point, an unstable black hole with larger event horizon
radius than critical radius, larger mass and lower temperature
changes to a stable phase with smaller radius, lower mass and
higher temperature because the branch with large radius has
negative heat capacity at constant potential but the other branch
with smaller radius has positive one. Then we analysis the
Ehrenfest's equations to prove that a second order phase
transition can happen for Bardeen black holes. The explicit
expressions of charge growth coefficient and isothermal
compressibility also reveal discontinuities at the same critical
point during the phase transition.

\vspace{1cm}
\noindent \textbf{Acknowledge}

This work is supported by NSFC No. 10875043 and is partly
supported by the Shanghai Research Foundation No. 07dz22020.

\newpage
\begin{figure}
\setlength{\belowcaptionskip}{10pt} \centering
\includegraphics[width=15cm]{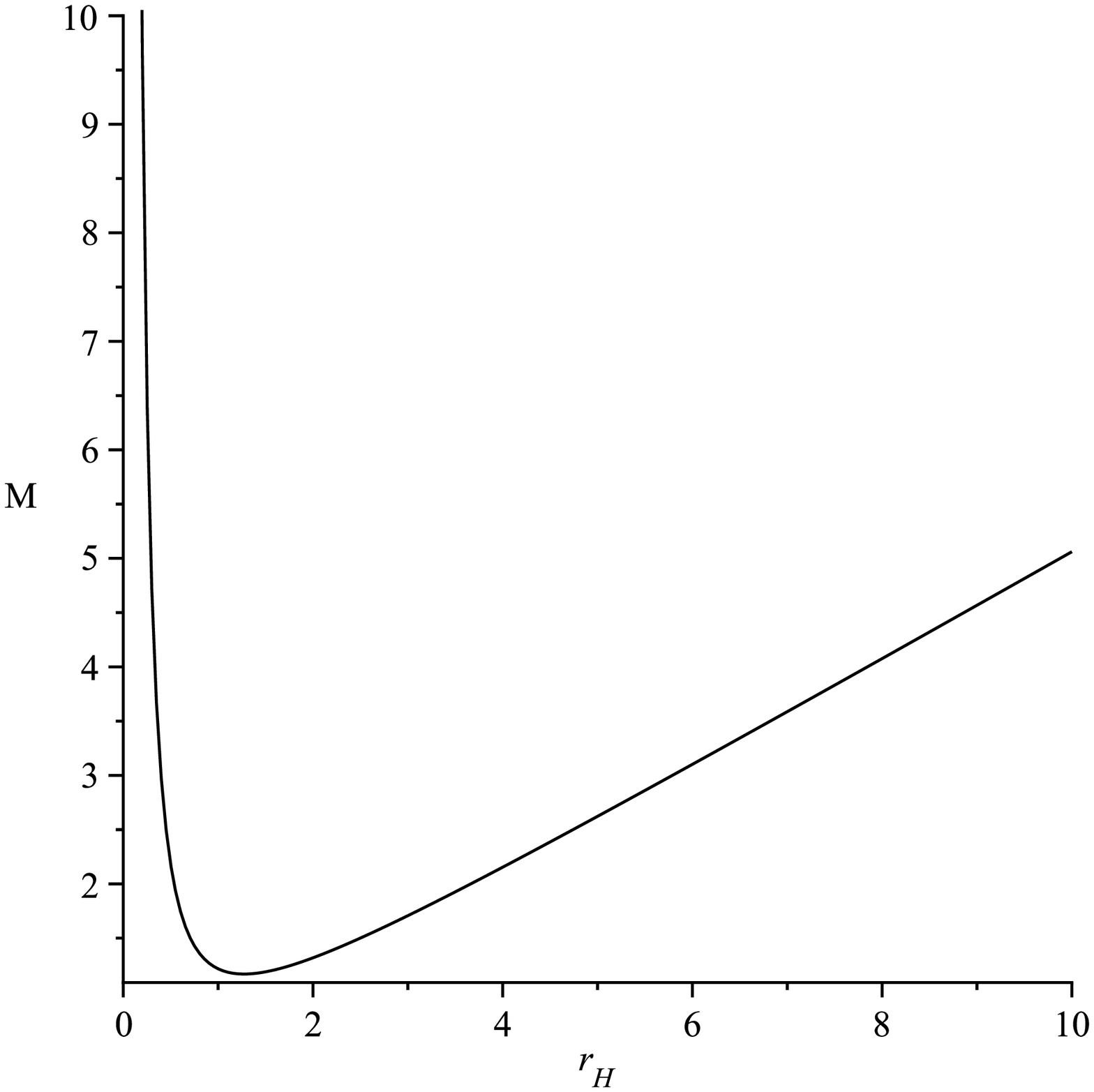}
\caption{The figure shows the mass of the magenetic monopole
as a function of event horizon radius for $q=0.9$.}
\end{figure}

\newpage
\begin{figure}
\setlength{\belowcaptionskip}{10pt} \centering
\includegraphics[width=15cm]{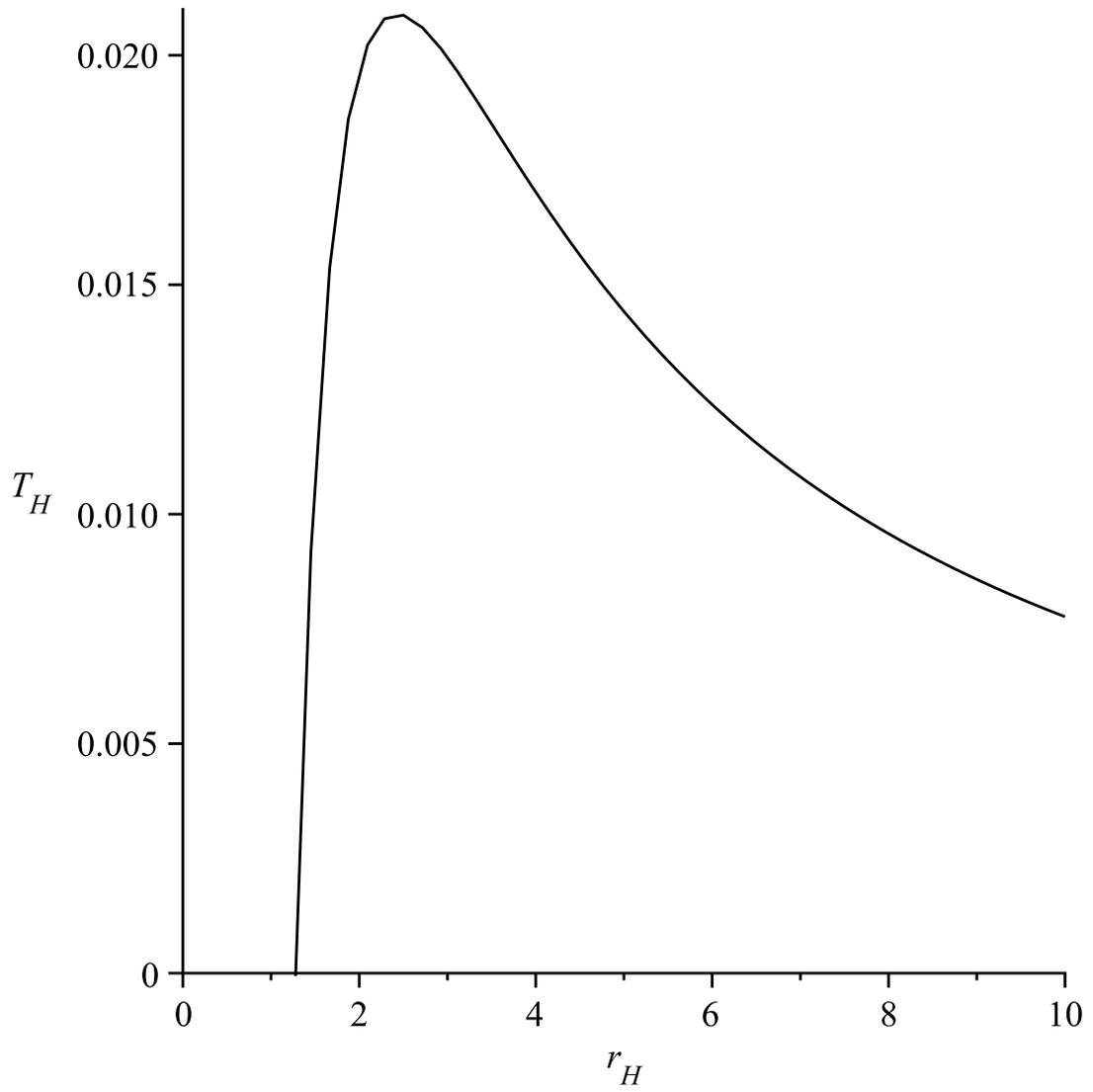}
\caption{The figure shows the temperature $T_{H}$
as a function of the event horizon radius $r_{H}$ for $q=0.9$.}
\end{figure}

\newpage
\begin{figure}
\setlength{\belowcaptionskip}{10pt} \centering
  \includegraphics[width=15cm]{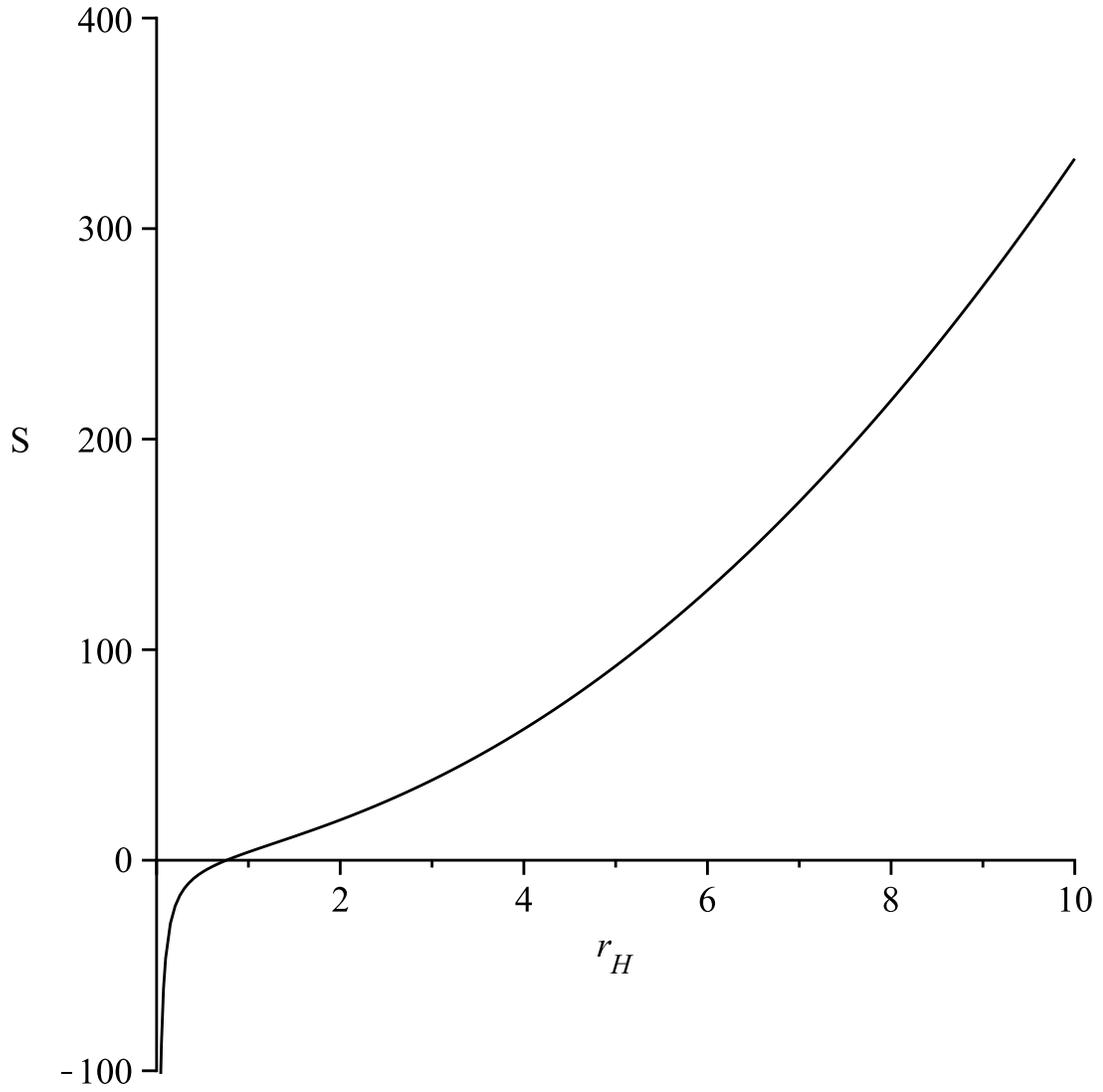}
  \caption{The relation between the entropy $S$
  and event horizon radius $r_{H}$ for $q=0.9$.}
\end{figure}

\newpage
\begin{figure}
\setlength{\belowcaptionskip}{10pt} \centering
  \includegraphics[width=15cm]{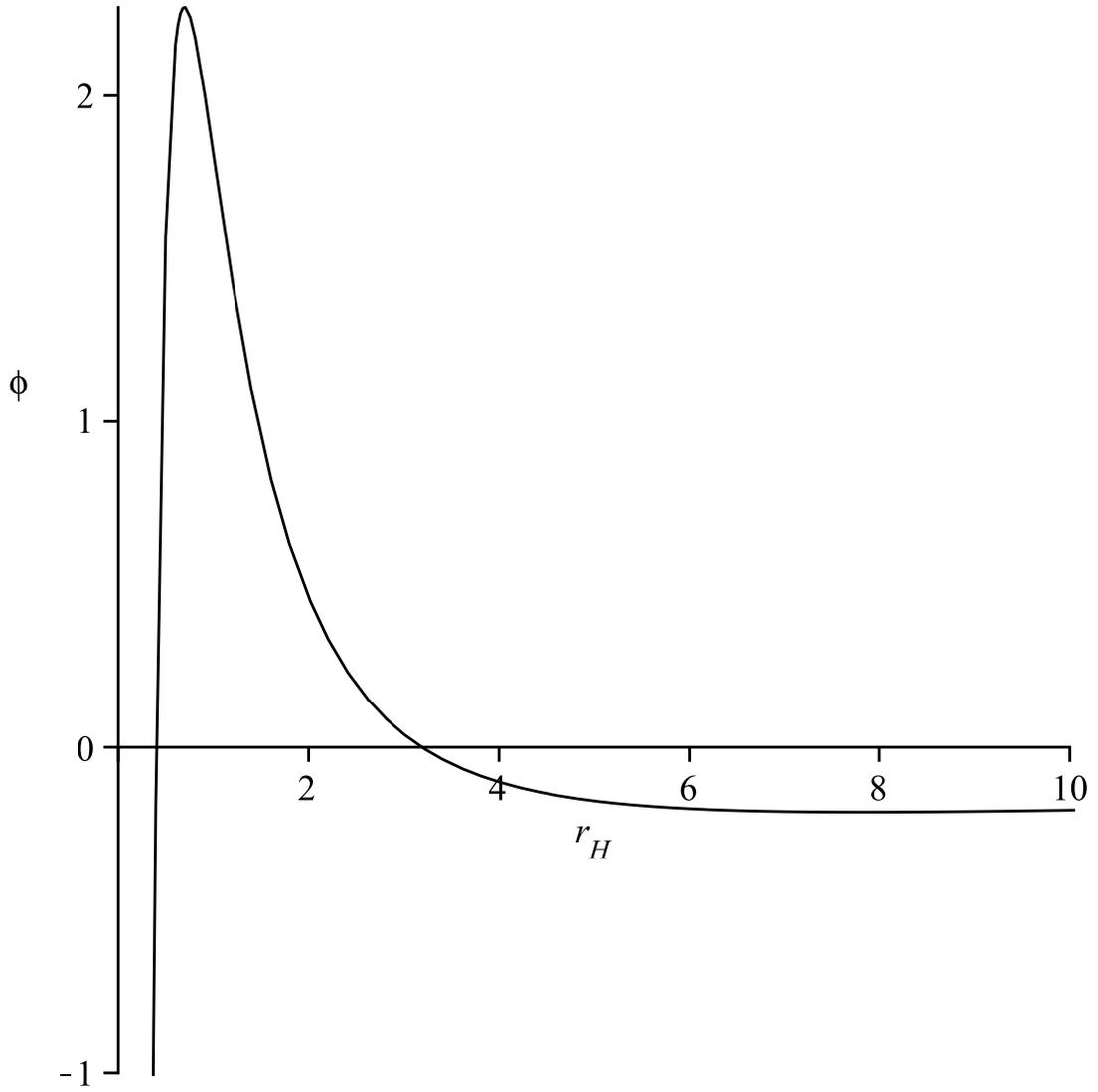}
  \caption{The relation between the electric potential $\phi$
  and event horizon radius $r_{H}$ for $q=0.9$.}
\end{figure}

\newpage
\begin{figure}
\setlength{\belowcaptionskip}{10pt} \centering
  \includegraphics[width=15cm]{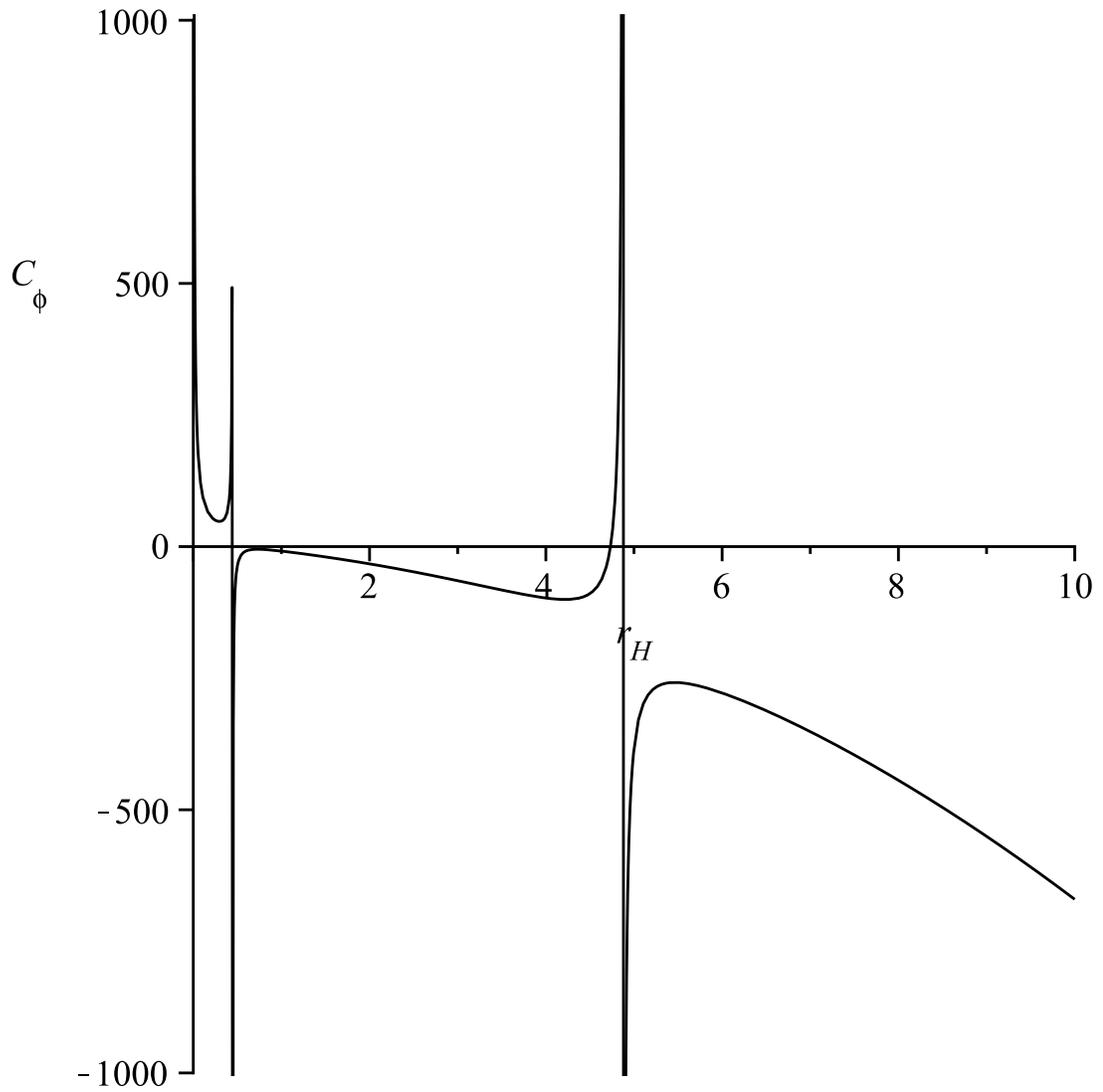}
  \caption{The behavior of the heat capacity at constant potential
  with the event horizon radius for the Bardeen black hole
  with charge $q=0.9$.}
\end{figure}

\newpage
\begin{figure}
\setlength{\belowcaptionskip}{10pt} \centering
  \includegraphics[width=15cm]{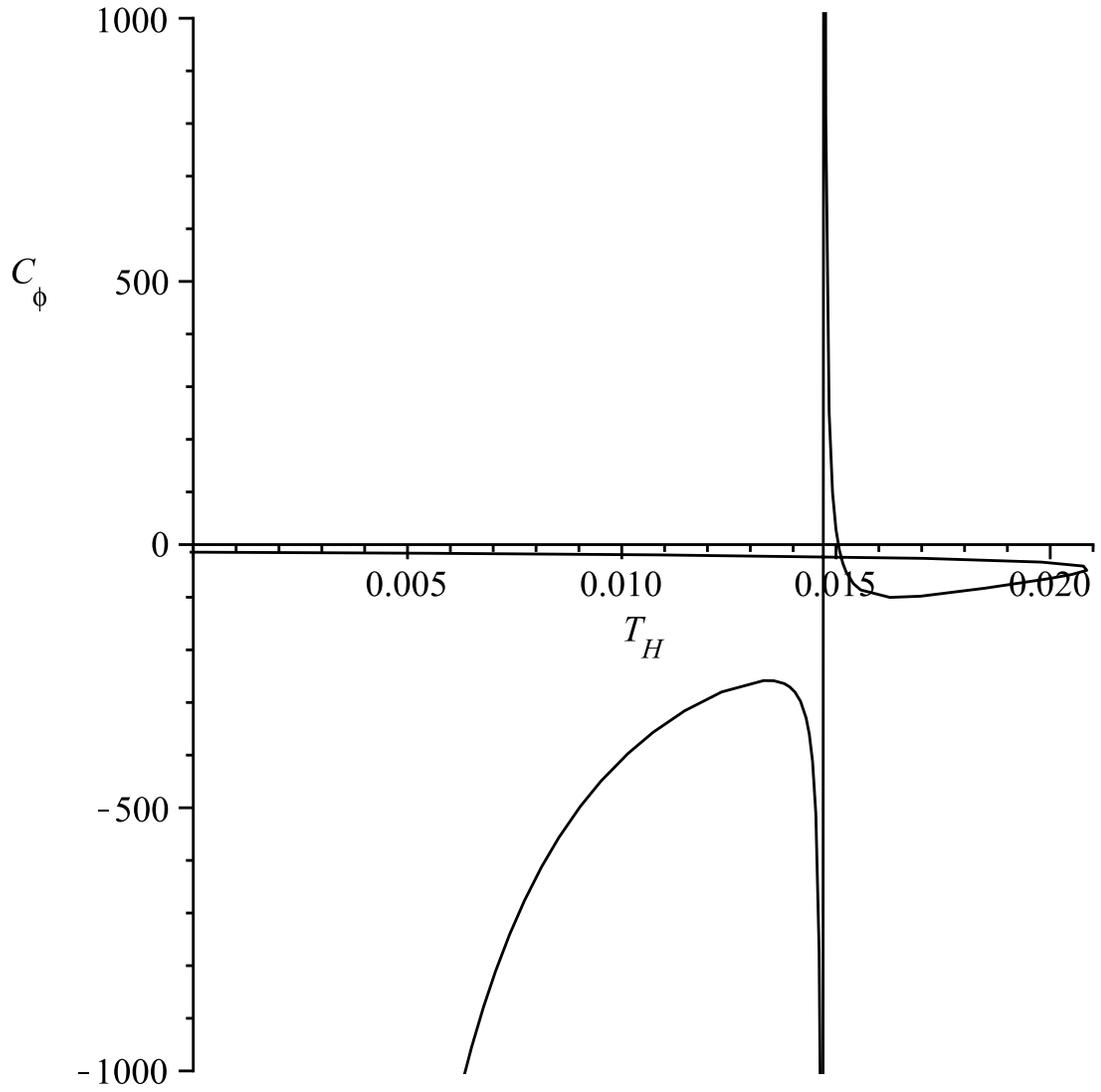}
  \caption{The behavior of the heat capacity at constant potential
  with the temperature for the Bardeen black hole with charge $q=0.9$.}
\end{figure}

\newpage
\begin{figure}
\setlength{\belowcaptionskip}{10pt} \centering
  \includegraphics[width=15cm]{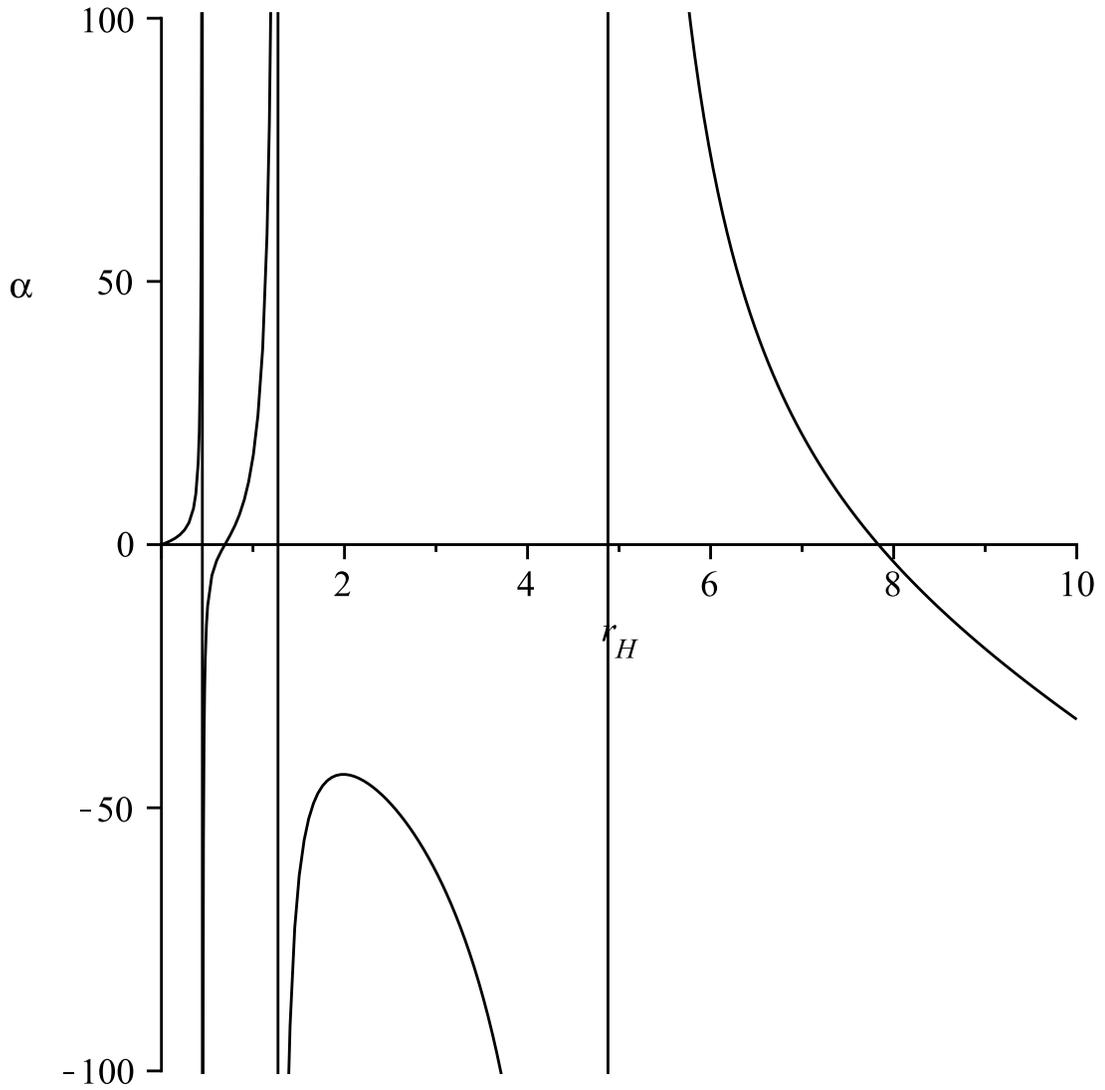}
  \caption{The behavior of charge growth coefficient
  with event horizon radius of the Bardeen black hole for $q=0.9$.}
\end{figure}

\newpage
\begin{figure}
\setlength{\belowcaptionskip}{10pt} \centering
  \includegraphics[width=15cm]{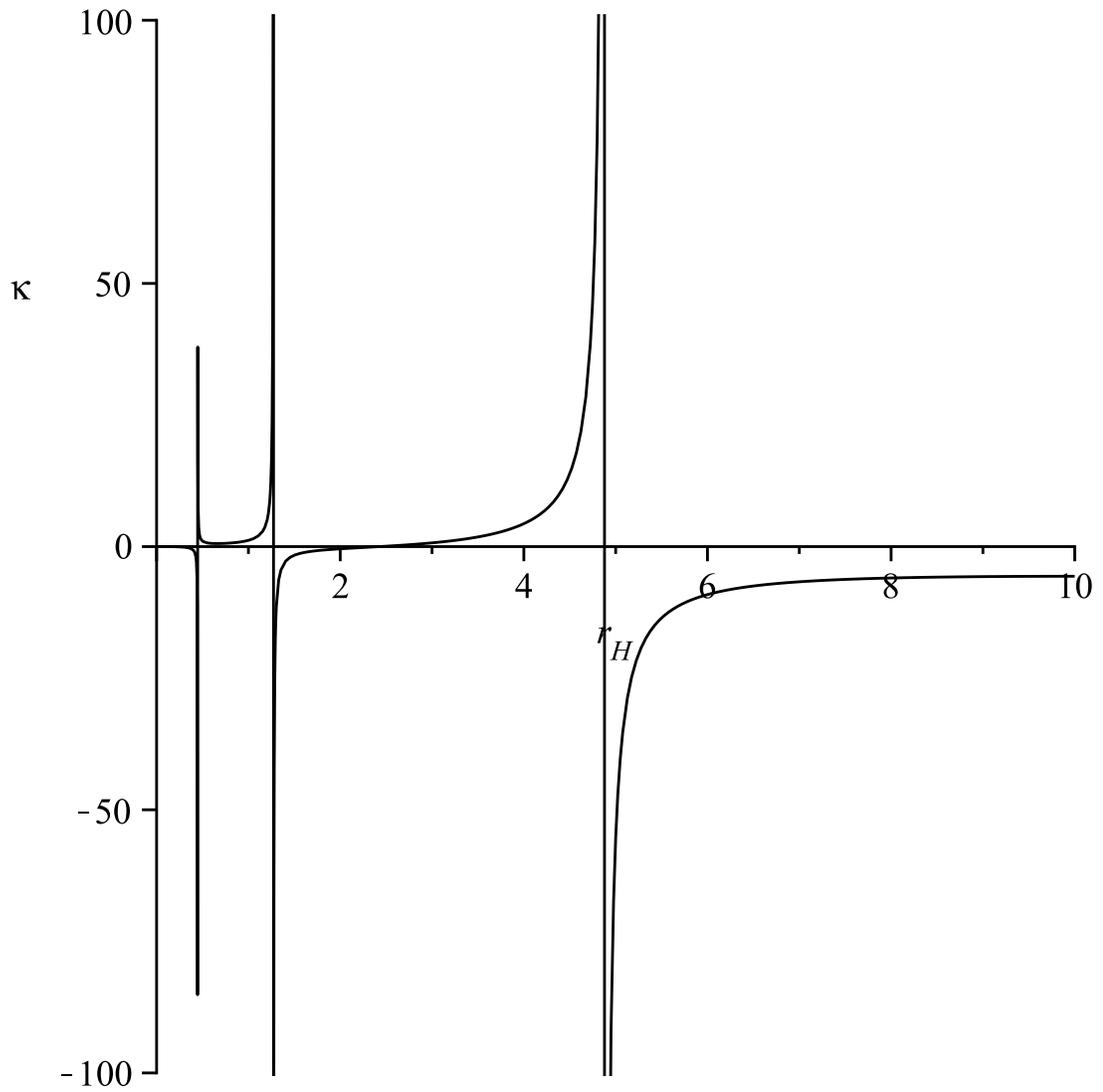}
  \caption{The behavior of the isothermal compressibility
  with event horizon radius of the Bardeen black hole for $q=0.9$.}
\end{figure}

\end{document}